\documentclass[iop]{emulateapj}

\usepackage{color}

\shorttitle{X-rays from I Zw 18}
\shortauthors{Kaaret \& Feng}

\begin{document}

\title{A State Transition of the Luminous X-ray Binary in the Low-Metallicity Blue Compact Dwarf Galaxy I Zw 18}

\author{Philip Kaaret\altaffilmark{1} and Hua Feng\altaffilmark{2}}

\altaffiltext{1}{Department of Physics and Astronomy, University of Iowa, Van Allen Hall, Iowa City, IA 52242, USA}

\altaffiltext{2}{Department of Engineering Physics and Center for Astrophysics, Tsinghua University, Beijing 100084, China}

\begin{abstract}

We present a measurement of the X-ray spectrum of the luminous X-ray binary in I Zw 18, the blue compact dwarf galaxy with the lowest known metallicity.  We find the highest flux yet observed, corresponding to an intrinsic luminosity near $1 \times 10^{40} \rm \, erg \, s^{-1}$ establishing it as an ultraluminous X-ray source (ULX).  The energy spectrum is dominated by disk emission with a weak or absent Compton component and there is no significant timing noise; both are indicative of the thermal state of stellar-mass black hole X-ray binaries and inconsistent with the Compton-dominated state typical of most ULX spectra.  A previous measurement of the X-ray spectrum shows a harder spectrum that is well described by a powerlaw.  Thus, the binary appears to exhibit spectral states similar to those observed from stellar-mass black hole binaries.  If the hard state occurs in the range of luminosities found for the hard state in stellar-mass black hole binaries, then the black hole mass must be at least $85 M_{\odot}$.  Spectral fitting of the thermal state shows that disk luminosities for which thin disk models are expected to be valid are produced only for relatively high disk inclinations, $\gtrsim 60\arcdeg$, and rapid black hole spins.  We find $a_* > 0.98$ and $M > 154 \, M_{\odot}$ for a disk inclination of $60\arcdeg$.  Higher inclinations produce higher masses and somewhat lower spins.

\end{abstract}


\section{Introduction}

The early universe was deficient in heavy elements.  Thus, understanding the metallicity dependence of X-ray emission and compact object formation is important for our understanding of the early universe, particularly the mechanisms of reionization \citep{Madau04,Mirabel11} and the effect of feedback from compact objects produced by star formation on the dynamics of early galaxies \citep{Power09}.  

Recent models of the formation and evolution of stars and binaries with low metallicity suggest that their X-ray emission is significantly enhanced \citep{Dray06,Dray07,Linden10}.  These trends are supported by observations.  \citet{Mapelli10} demonstrated an increase in the number of X-ray bright HMXBs (for a given SFR) for moderately low metallicity galaxies.  \citet{Kaaret11} found a suggestion of enhanced X-ray luminosity (for a given SFR) in a sample of very-metal-poor blue compact dwarf galaxies.  

It has also been suggested that low metallicity environments may lead to the formation of unusually massive compact objects.  Solar metallicity stars are expected to lose much of their mass via winds at the ends of their lives and produce remnants with masses less than about 20~$M_{\odot}$.  Reduced metallicity decreases the effect of winds and leads to more massive remnants \citep{Belczynski10}.  This has been suggested to explain the increased occurrence of X-ray binaries with very high luminosities, the ultraluminous X-ray sources \citep{Kaaret01,Feng11}, in low metallicity environments \citep{Pakull03,Zampieri04,Mapelli09}.

Blue compact dwarf galaxies (BCDs) are physically small galaxies with blue optical colors and low metallicities \citep{Kunth00,Wu06}.  They are the best local analogs to early galaxies yet identified.   I Zw 18 is a nearby BCD with the lowest known metallicity and the highest X-ray luminosity (Kaaret et al.\ 2011).  The X-ray emission from I Zw 18 is dominated by a single X-ray binary and observations reported to date have shown a maximum luminosity below the Eddington luminosity of a 20~$M_{\odot}$ compact object.

Here, we report on a measurement of the X-ray spectrum of I Zw 18 that reveals the highest flux yet observed.  The spectrum shows distinct curvature indicative of a black hole X-ray binary in the thermal state.  We present our results on the X-ray spectrum in \S 2 and discuss their implications for the nature of compact object in \S 3.

\begin{deluxetable*}{lccccc}
\tabletypesize{\scriptsize}
\tablecaption{X-ray Spectral Fits\label{xray_spec}}
\tablewidth{0pt}
\tablehead{Model & $\chi^2$/DoF & $L_X$ & $N_H$      & $kT/E_c$ & $\Gamma$ \\
                 &              & (10$^{40}$~erg~s$^{-1}$)
                                        & (10$^{21}$~cm$^{-2}$) 
                                                     &  (keV)   &  }
\startdata
Powerlaw         & 220.8/151   & 3.0  & 3.8$\pm$0.5 &                     & 2.01$\pm$0.06 \\
Diskbb           & 172.0/151   & 1.1  & 0.7$\pm$0.3 & 1.05$\pm$0.05       &               \\
Cutoff powerlaw  & 156.7/150   & 1.2  & 1.4$\pm$0.6 & 2.1$\pm$0.6         & 0.80$\pm$0.26 \\
Diskbb+powerlaw  & 157.3/150   & 1.3  & 1.4$\pm$0.6 & 1.04$\pm$0.07       & 2             \\
Simpl*kerrbb     & 153.4/150   & 1.2  & 1.2$\pm$0.4 &                     & 2             \\
\enddata
\tablecomments{The table includes: the model name, goodness of fit ($\chi^2$) and degrees of freedom (DoF), the intrinsic luminosity in the 0.3--10~keV band assuming a distance of 18.2~Mpc \citep{Aloisi07} and isotropic emission, absorption column density for the component with abundances fixed to $Z/Z_{\odot} = 0.019$, the disk temperature at the inner edge ($kT$) or the powerlaw cutoff energy ($E_c$), and the photon index ($\Gamma$).}
\end{deluxetable*}

\begin{figure}[tb]
\centerline{\includegraphics[width=2.75in,angle=0]{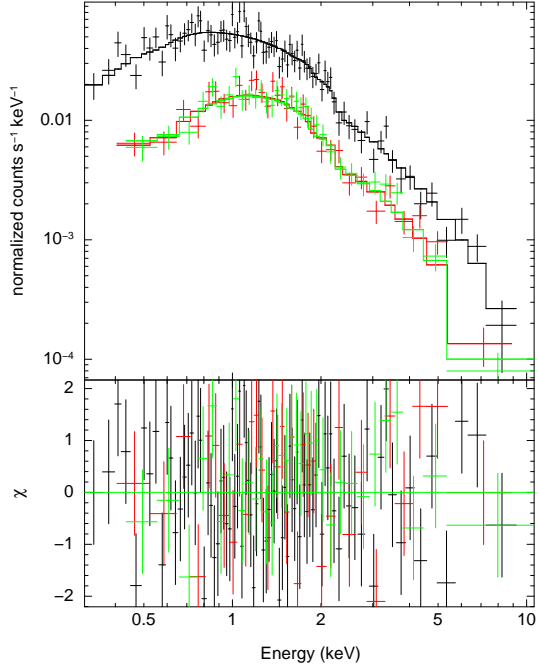}}
\caption{X-ray spectrum of I Zw 18.  The solid, stepped curves show the best fit to the simpl*kerrbb model with absorption as described in the text.  There are three curves and three sets of data points.  The upper curve/data is for the pn.  The lower two curves/data are for the MOS1 and MOS2.}
\label{izw18spec_kerrbb_simpl} \end{figure}

\section{Observations and Analysis}

I Zw 18 was observed by the X-ray Multi-Mirror Mission (XMM-Newton) on 2002 April 10 for 32.8~ks and on 2002 April 16 for 28.9~ks.  We reduced the data according to the standard procedures for imaging spectroscopy with the European Photon Imaging Camera (EPIC).  Background flaring reduced the useful exposure for the first observation for the pn camera to 24.1~ks and for the two MOS cameras to 30.9~ks.  Flaring was much worse for the second exposure and reduced the useful exposure for the pn to 5.3~ks.  Thus, we chose to further analyze only the first observation.  

We extracted spectra using circular source regions with $30\arcsec$ radius centered on the X-ray source apparent in each image.  This extraction region is larger than the extent of I Zw 18 and therefore includes all X-ray emission from the galaxy.  Background subtraction was performed using a circular background region with a radius of $60\arcsec$ located on the same CCD chip.  Together the 3 spectra contain about 3940 net counts.   Spectral response files suitable for point source analysis were calculated using the most recent calibrations.  The spectra were grouped to have a minimum of 16 counts per bin.  We fitted the X-ray spectra using the xspec software package \citep{Arnaud96}.  We used the 0.3-10~keV energy range for fitting.

The Chandra X-ray Observatory observed I Zw 18 for 41~ks on 2000 February 8 and imaging and spectroscopic results were reported by \citet{Bomans02} and \citet{Thuan04}.  The imaging shows that the X-ray emission was dominated by a single point source with at least 96\% of the flux.  Both groups found good fits with an absorbed power-law model, but residuals near 0.65~keV that they ascribed to an O{\sc III} hydrogen-like line.  \citet{Thuan04} report an observed flux of $7.2 \times 10^{-14} \rm \, erg \, cm^{-2} \, s^{-1}$ in the 0.5-10~keV band, a photon index of $2.01 \pm 0.16$, and an absorption column density, $N_H = (1.44 \pm 0.38) \times 10^{21}$~cm$^{-2}$.

We fitted the XMM spectra using an absorbed powerlaw model and found an observed flux of $(2.78 \pm 0.12)\times 10^{-13} \rm \, erg \, cm^{-2} \, s^{-1}$ in the 0.5-10~keV band, a photon index of $2.31 \pm 0.09$, and an absorption column density, $N_H = (2.7 \pm 0.3) \times 10^{21}$~cm$^{-2}$.  Allowing the normalization to vary between the 3 detectors did not improve the fit, so we used the same normalization for all three detectors during all subsequent fitting.  The marked increase in flux from the Chandra to the XMM observation shows that the emission is variable and likely predominantly due to a single X-ray binary.  Thus, we consider only spectral models appropriate for X-ray binaries.  Since the XMM extraction region includes the whole galaxy, it is possible that a second source, other than the one detected with Chandra, contributes a significant fraction of the flux observed in the XMM observation.  Due to the low numbers of X-ray sources detected in similar galaxies \citep{Kaaret11}, this is unlikely, but could be tested with a new Chandra observation.

The absorption column density required for these spectral fits is well above the total Galactic H{\sc i} column density towards I Zw 18 of $N_H = 2.5 \times 10^{20}$~cm$^{-2}$.  Thus, most of the absorbing material likely resides in I Zw 18.  The metallicity of I Zw 18 is measured via optical spectroscopy of H{\sc ii} regions to be $Z/Z_{\odot} = 0.019$ \citep{Izotov99}, where we have adopted a solar oxygen abundance of 12+log(O/H) = 8.9, while the abundances in the neutral interstellar medium are several times lower \citep{Aloisi03}.  The X-ray binary imaged with Chandra lies in a star formation region and the H{\sc ii} region abundance is appropriate.  The X-ray absorption model used above and also by \citet{Bomans02} and \citet{Thuan04} assumes solar abundance and is likely an incorrect description of the true energy dependence of the absorption.  In the fits below, we use two absorption components: one with solar abundances with a column density fixed to the Galactic H{\sc i} column towards I Zw 18 (TBabs in xspec) and a second with a variable column density (TBvarabs), abundances fixed to $Z/Z_{\odot} = 0.019$, and redshift fixed to 0.00254.

The results of fitting to various models are shown in Table~\ref{xray_spec}.  The only model which is excluded is the simple powerlaw, while the diskbb model is marginally excluded.  A model consisting of the sum of these two components provides a good fit.  The three best fitting models are statistically indistinguishable.  They have an observed flux of $2.7 \times 10^{-13} \rm \, erg \, cm^{-2} \, s^{-1}$ in the 0.3--10~keV band and the absorption column density within I Zw 18 is in the range 1.2--1.4$\times 10^{21} \rm \, cm^{-2}$.  \citet{Lelli12} mapped H{\sc i} in I Zw 18 via the 21~cm line at 2$\arcsec$ resolution and their maps show a total column density near $6 \times 10^{21} \rm \, cm^{-2}$ at the X-ray source position.  The lower $N_H$ values from the X-ray spectral fits are reasonable if the X-ray binary lies closer than the midplane of the galaxy or if the metallicity of the H{\sc i} gas is lower than that of the H{\sc ii} regions, suppressing X-ray absorption.  We note that none of these model fits requires an emission line at 0.65~keV and we suggest that the line was due to an inappropriate choice of absorption model.

The cutoff powerlaw model is empirical, but provides a simple analytical form, $f(E) = E^{-\Gamma} \exp(-E/E_c)$.  The improvement in the fit going from the powerlaw to the cutoff powerlaw indicates spectral curvature at high energies.

The sum of a powerlaw and multicolor disk blackbody (diskbb) is often used to model the spectral of Galactic black hole X-ray binaries \citep{Remillard06}.  In our fitting, the photon index ($\Gamma$) was not well constrained, so we fixed $\Gamma = 2$.  The powerlaw normalization was $2.3 \times 10^{-5}$ and the diskbb normalization was $7.8 \times 10^{-3}$.  The disk component produces 76\% of the flux in the 2--10~keV band.

\begin{figure}[tb]
\centerline{\includegraphics[height=3.25in,angle=0]{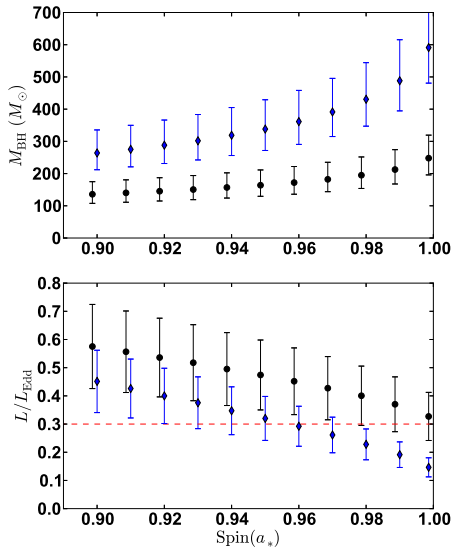}}
\caption{Dependence of black hole mass ($M_{BH}$, upper panel) and Eddington ratio ($L/L_{Edd}$, lower panel) on black hole spin ($a_{*}$) for disk inclinations of $i = 60\arcdeg$ (circles) and $i = 75\arcdeg$ (diamonds).  The dashed line in the lower panel shows $L/L_{Edd} = 0.3$.}\label{kerrbb} 
\end{figure}

The kerrbb model represents a sophisticated model of a thin accretion disk around a Kerr black hole including all relativistic effects and also self-irradiation of the disk due to light deflection \citep{Li05}.  The simpl model adds a Comptonization component calculated by convolution of the disk spectrum \citep{Steiner09}.  We fixed the source distance to 18.2~Mpc \citep{Aloisi07}, the spectral hardening factor to 1.7 \citep{Shimura95}, the torque at the inner disk boundary to zero, and included self irradiation and limb darkening in the model.  The fit was insensitive to the black hole spin parameter ($a_{*}$), inclination ($i$), and Comptonization photon index ($\Gamma$).  The best fitted spectrum for $a_{*} = 0.9986$, $i=60\arcdeg$, $\Gamma = 2.0$ is shown in Fig.~\ref{izw18spec_kerrbb_simpl}.  The fitted parameters not included in Table~\ref{xray_spec} were: black hole mass, $M_{BH} = 249^{+71}_{-52} M_{\odot}$, mass accretion rate, $\dot{M} = (35 \pm 3) \times 10^{18} \rm \, g \, s^{-1}$, and the Compton scattered fraction, $0.27 \pm 0.13$.  We also performed fits with $\Gamma = 2.6$, to cover the range typically found in the thermal state, and none of the fit parameters changed significantly.

We performed fits for spins over a range from $a_{*} = 0.9$ to the maximum achievable spin $a_{*} = 0.9986$ \citep{Li05} for inclinations of $60\arcdeg$ and $75\arcdeg$.  Lower inclinations and lower spins produce Eddington ratios, $L/L_{Edd} > 0.3$, inconsistent with application of the kerrbb model.  The spin was kept fixed for each individual fit.  The $\chi^2$ varied by less than 0.5 over this range.  The Comptonized fraction did not vary significantly, with best fitted values in the range 0.26--0.33.  The absorption column also did not vary significantly, with best fitted values in the range 1.0--1.2 $\times 10^{21}$~cm$^{-2}$.  The best fitted black hole mass ($M_{\rm BH}$) and accretion rate vary with both $a_{*}$ and $i$.  Fig.~\ref{kerrbb} show the variation in $M_{\rm BH}$ and the ratio of disk luminosity to the Eddington luminosity ($L/L_{Edd}$).

For comparison with \citet{Gladstone09}, we fitted the XMM data in the 2--10~keV band with a powerlaw and a broken powerlaw, each without absorption.  We note that our spectrum contains fewer counts, $\sim 4000$, than the $\gtrsim 10,000$ required by \citet{Gladstone09}for inclusion in their sample.  The powerlaw provides an adequate fit with $\chi^2/\rm DoF = 51.7/43$ and a photon index $\Gamma = 2.40 \pm 0.16$.  The broken powerlaw provides only a slight improvement, $\chi^2/\rm DoF = 48.4/41$ for an F-test value of 0.25.  The fit is insensitive to the low energy photon index, so we fixed it to 1.4 (one less than $\Gamma$ for the simple powerlaw).  The break energy is then between the lower end of the fitting range, 2.0~keV, and $2.9$~keV and the high energy photon index is $2.59 \pm 0.25$.  We note that fitting the full energy range to a broken powerlaw model with absorption leads to a break energy of $1.9 \pm 0.3$~keV.  We also fitted the data in the full energy range with the sum of a disk model (diskpn) and a Comptonization model (comptt) with absorption.  We tied the Comptonization photon input temperature to the disk temperature and used an inner disk radius of 6$R_G$.  The model produced a reasonable fit with $\chi^2/\rm DoF = 152.6/148$ and a disk temperature $kT = 0.5_{-0.2}^{+0.5}$~keV.  However, the other parameters were very poorly constrained.  Only lower bounds were obtained for the Compton optical depth, $\tau > 0.05$, and the plasma temperature, $kT_{e} > 1.14 \rm \, keV$.  \citet{Gladstone09} found significant improvement (F-test significance level of $>$99\%) of the best fitted model relative to a hot corona with $kT_e = 50$~keV for the spectra of sources identified as in the ultraluminous state.  Making the same comparison, we find $\Delta \chi^2= 0.8$ (DoF = 148) and an F-test significance level of 62\%.  Thus, the low temperature Compton component does not produce a statistically significant improvement in the fit, as required for the ultraluminous state.

\begin{figure}[tb]
\centerline{\includegraphics[width=3.25in,angle=0]{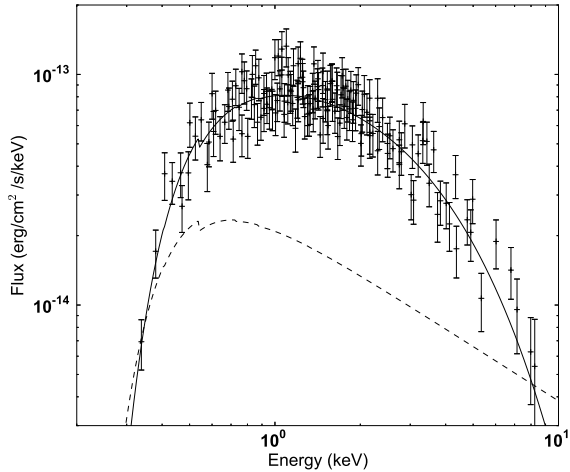}}
\caption{X-ray spectra of I Zw 18 at low and high flux levels.  The points and the solid curve show the XMM observation fitted with the cutoff powerlaw model.  The dashed curve shows the powerlaw model fitted to the Chandra data.}
\label{izw18spec_highlow} \end{figure}

We reduced the Chandra data previously analyzed by \citet{Bomans02} and \citet{Thuan04} and found a net 490 counts.  Fitting with an absorbed powerlaw, we found results consistent within errors with those of \citet{Thuan04}.  For comparison with the XMM-Newton results reported above, we also fitted the Chandra data with models with two absorption components, a Galactic component with solar abundances and $N_H = 2.5 \times 10^{20}$~cm$^{-2}$ and a component for I Zw 18 with $Z/Z_{\odot} = 0.019$ with variable $N_H$.  Fitting with a powerlaw, we found a good fit with $\chi^2/\rm DoF = 16.2/25$, $\Gamma = 1.78 \pm 0.21$, $N_H = 9^{+10}_{-8} \times 10^{20} \rm \, cm^{-2}$, and a flux of $8.3 \times 10^{-14} \rm \, erg \, cm^{-2} \, s^{-1}$ in the 0.3-10~keV band.  Fitting with a cutoff powerlaw produced an essentially identical fit with a best fitted cutoff energy of 500~keV and a 90\% confidence lower bound on the cutoff energy of 4.7~keV.  Fitting with a diskbb model produced a poor fit with $\chi^2/\rm DoF = 36.7/25$.   Thus, we conclude there is no evidence for spectral curvature at high energies in the Chandra spectrum.

We examined the timing properties of the XMM pn data.  The 0.3--10~keV light curve binned in 450.8~s intervals appears constant, with $\chi^2/\rm DoF = 55.8/54$.  The rms power integrated over 0.01-1~Hz is less than 5\% at 99\% confidence.  Due to the low number of counts, the Chandra data do not allow us to place useful constraints on the timing properties.

\section{Discussion}

The XMM-Newton observations described here show the highest flux yet observed from I Zw 18, corresponding to an intrinsic luminosity near $1 \times 10^{40} \rm \, erg \, s^{-1}$ and establishing the X-ray binary in I Zw 18 as an ultraluminous X-ray source (ULX).

The spectral shapes of the XMM versus Chandra data are different with the high flux (XMM) spectrum showing distinct curvature at high energies while the low flux (Chandra) spectrum shows no evidence of curvature, see Fig.~\ref{izw18spec_highlow}.  This is consistent with the state transitions seen in stellar-mass black hole X-ray binaries, specifically the transition between the hard X-ray spectral state and the thermal state.  The lack of timing noise and low fraction ($<30\%$) of powerlaw flux in the diskbb plus powerlaw model reported in Table~\ref{xray_spec} for the high flux state are consistent with its identification as the thermal state \citep{Remillard06}.


\citet{Gladstone09} has suggested the existence of an `ultraluminous' state based on curvature at high energies in the spectra of several ULXs.  Broken powerlaw fits to the XMM spectra of I Zw 18 produce break energies well below any of those reported for the ultraluminous state.  Also, a spectral model with a low temperature Compton component ($kT_e = $ 1--3~keV) produces no statistically significant improvement in the fit relative to a model with a hot Compton component ($kT_e = $50~keV).  These results suggest that the XMM spectrum is inconsistent with the criteria established for the ultraluminous state.  However, we caution that the spectrum contains fewer counts than those used to establish the properties of the ultraluminous state.  A higher quality spectrum with at least $3\times$ the number of counts would be needed to draw definitive conclusions.

State transitions have been reported in ULXs that have been modeled and interpreted as being unlike those of stellar-mass black hole X-ray binaries.  \citep{Pintore12} analyzed multiple XMM observations of the two ULXs in NGC 1313 and classified their spectra into two states: the `thick corona' state and the `very thick corona' state.  The spectra in both states are dominated by a coronae with $kT_e$ in the range 1--6~keV.  Other authors have reported on spectral evolution in ULXs with the common characteristic that a Compton component with $kT_e > $1~keV, contributes a major and usually dominant fraction of the flux \citep{Feng09,Kajava12}.  In the XMM spectra of I Zw 18, a Compton component modeled as a powerlaw produces a low fraction of the X-ray flux and a Compton component with a low temperature ($kT_e = $ 1--3~keV) produces no statistically significant improvement in the fit.  Thus, the spectral state transitions in most ULXs appear different from that seen in I Zw 18.

While rare, the thermal state and transitions between the hard state and the thermal state are sometimes found in ULXs.  The strongest case for such a transition is for M82 X-1 which shows both a spectral transition and a simultaneous change in the X-ray timing properties of the source as seen in stellar-mass black hole X-ray binaries \citep{Feng10}.  Spectral evidence for the hard/thermal state transition has also been presented for the extremely luminous X-ray binary in ESO 243-49 \citep{Servillat11} and for M82 X37.8+54 \citep{Jin10}.  Spectral evidence for the thermal state was found for NGC 247 X-1 \citep{Jin11} and at lower luminosities in CXOM31 J004253.1+411422 \citep{Middleton12}.

Detection of a state transition similar to those seen in stellar-mass black hole X-ray binaries strengthens interpretation of the hard spectrum seen in the Chandra observation as evidence that the source was in the hard state.  The maximum luminosities observed from stellar-mass black holes in the hard state are less than $0.3 L_{\rm Edd}$ \citep{Rodriguez03,Zdziarski04,Yuan07,Miyakawa08}.  The luminosity of the binary in I Zw 18 while in the hard state was $3.3 \times 10^{39} \rm \, erg \, s^{-1}$.  This corresponds to $L < 0.3 L_{\rm Edd}$ only if $M_{BH} > 85 M_{\odot}$.  

In the thermal state, the inner radius of the accretion disk is set by the mass and spin of the black hole.  As described above, we found that the spectral fits with simpl*kerrbb model were insensitive to the black hole spin ($a_{*}$) and inclination ($i$), thus, we considered a range of values for $a_{*}$ and $i$.  For $i <60\arcdeg$, the source is super-Eddington for all spins.  The accretion disk is expected to be geometrically thin, and thus the kerrbb model is expected to be valid, only for $L/L_{Edd} < 0.3$ \citep{McClintock06}, so we considered only fits with $i > 60\arcdeg$.  Fig.~\ref{kerrbb} shows the dependence of black hole mass ($M_{BH}$) and disk luminosity relative to Eddington ($L/L_{Edd}$) on $a_{*}$ for $i = 60\arcdeg$.  We find that $L/L_{Edd} \le 0.3$ (within errors) for $a_{*} > 0.98$ with $M_{BH} = 195_{-41}^{+57} M_{\odot}$.  For higher spins, $L/L_{Edd}$ decreases and $M_{BH}$ increases.  For $i = 75\arcdeg$, we find that $L/L_{Edd} \le 0.3$ for $a_{*} > 0.92$ with $M_{BH} = 288_{-57}^{+78} M_{\odot}$.  Again, $L/L_{Edd}$ decreases and $M_{BH}$ increases for higher spins.

In conclusion, the X-ray spectrum from the binary in I Zw 18 while in the high flux state can be interpreted in terms of a non-rotating black hole with super-Eddington emission and a mass similar to that of the most massive known stellar-mass black hole \citep{Silverman08,Prestwich07}.  However, if the nature of the state transition seen from the source is similar to those seen from stellar-mass black hole X-ray binaries, including the luminosity threshold for the transition, then the compact object is likely a near maximally-rotating black hole with an unusually high mass $M_{BH} > 85 M_{\odot}$.  This is close to the maximum mass of $75 M_{\odot}$ predicted for black holes formed via stellar collapse in a low metallicity ($Z/Z_{\odot} = 0.019$) environment \citep{Belczynski10}.  Modeling of the thermal state spectrum suggests a higher mass, $M_{BH} >= 154 M_{\odot}$, in the IMBH range.  We note that the X-ray source is coincident with a massive star cluster, sites suggested as potential places of original for IMBHs \citep{Portegies04}, particularly in low-metallicity environments \citep{Mapelli13}.  Due to their similarities to early galaxies, identification of an IMBH in a BCD would be of particular interest for our understanding of the formation of early galaxies and supermassive black holes \citep{Reines11}.

\acknowledgments

We thank the referee for comments that improved the paper.  This research has made use of data obtained from the Chandra Data Archive and software provided by the Chandra X-ray Center (CXC) in the application packages CIAO, ChIPS, and Sherpa.

\end{document}